\documentclass[preprint,12pt,authoryear]{elsarticle}

\journal{Annals of Physics}

\usepackage{graphicx}
\usepackage[english]{babel}
\usepackage{bm}
\usepackage{amsmath,amsfonts}
\usepackage{amsbsy}
\usepackage[colorlinks=true,linkcolor=blue,urlcolor=blue,citecolor=blue,breaklinks=true]{hyperref}
\usepackage[utf8]{inputenc}

\usepackage{graphics}
\usepackage{subcaption}
\usepackage{xcolor}
\usepackage{bbm}
\usepackage[normalem]{ulem}
\usepackage{amssymb}
\usepackage{array}
\usepackage{amsmath}
\usepackage{graphicx}\usepackage{graphics}
\usepackage{dcolumn}
\usepackage{bm}\usepackage{varioref}
\begin{document}

\begin{frontmatter}

\title{Giant microwave absorption in $s$- and $d$- wave superconductors}

\author[1]{M. Smith}

\author[1,2,3]{A. V. Andreev}

\author[1] {B.Z. Spivak}
\address[1]{Department of Physics, University of Washington, Seattle, WA 98195,  USA}

\address[2]{Skolkovo  Institute of  Science  and  Technology,  Moscow,  143026,  Russia}
\address[3]{ L. D. Landau Institute for Theoretical Physics, Moscow, 119334 Russia}

\date{July 18, 2019}

\begin{abstract}
\,\,\,\,\,\,\,\,\,\,\,\,\,\,\,\,\,\,\,\,\,\,\,\,\,\,\,\,\,\,\,\,\,\,\,\,\,\,\,\,\,\,\,\,\,\,\,\,\,\,\,\,\,\,\,\,\,\,\,\,\,\,\,\,\,\,\,\, {\it  Dedicated to  Gerasim Eliashberg's 90-th birthday.  } \smallskip

We discuss a new mechanism of microwave absorption  in $s$- and $d$-wave superconductors, which arises in the presence of a \emph{dc} supercurrent in the system. It produces a contribution to the \emph{ac} conductivity that is proportional to the inelastic quasiparticle relaxation time. This contribution also determines the supercurrent dependence of the conductivity. It may significantly exceed
the conventional contribution because in typical superconductors the inelastic relaxation time
is  several orders of magnitude longer than the elastic one.
We show that the
 aforementioned contribution to the conductivity may be expressed in terms of the single particle density of states in superconductors in the presence of a  \emph{dc} supercurrent. Our results may enable determination of the inelastic relaxation time in superconductors from
microwave absorption measurements.
\end{abstract}

\end{frontmatter}

\section{Introduction}

In this article we discuss the theory of microwave absorption in superconductors. In linear response to the microwave field $\bm{E}(t)=  \bm{E}_\omega \cos (\omega t)$,  and in the limit of low frequencies $\omega$, the current density in a superconductor may be written as
\begin{equation}
\label{eq:current}
{\bf j}=\frac{e}{m} N_\mathrm{s}\, \bm{p}_\mathrm{s}+\sigma {\bf E}.
\end{equation}
Here $N_\mathrm{s}$ is the superfluid density, $e$ and $m$ are, respectively, the charge and the mass of the electron, and the superfluid momentum is defined by $\bm{p}_{s}= \frac{\hbar}{2} \bm{\nabla} \chi - \frac{e}{c}\bm{A}$, with $\chi$ being the  phase of the order parameter, and $\bm{A}$ the vector potential. The second term in Eq.~\eqref{eq:current}, characterized by the conductivity $\sigma$, represents the dissipative part of the current.

The  microwave absorption coefficient is controlled by the conductivity $\sigma$. The value of $\sigma$ is determined by the quasiparticle scattering processes in the superconductor, which are generally characterized by two relaxation times: elastic, $\tau_{\mathrm{el}}$, and inelastic, $\tau_{\mathrm{in}}$, ones.  In a typical situation, which we assume below, $ \tau_{\mathrm{in}} \gg \tau_{\mathrm{el}}$.    The theory of transport phenomena in conventional superconductors was developed long ago, see for example  ~\cite{mattis_theory_1958,schrieffer_theory_1964,larkin_nonlinear_1977,ovchinnikov_electromagnetic_1978,aronov_boltzmann-equation_1981}.
The conventional result is that the conductivity, and consequently the microwave absorption coefficient, are proportional to the elastic relaxation time $\tau_{\mathrm{el}}$.  For example, at temperatures $T$ near the critical temperature $T_{c}$, the conductivity of a superconductor is close to the normal metal Drude conductivity $\sigma_{\mathrm{D}} =e^2\nu_n D$.
Here  $D=v_{\mathrm F}^{2}\tau_{\mathrm{el}}/3$ is the diffusion coefficent, $\nu_{n}$ is the normal metal density of states at the Fermi level, and $v_{\mathrm F}$ is the Fermi velocity.

In this article we discuss another  contribution to
the conductivity, $\sigma_\mathrm{DB}$, that is proportional to the inelastic relaxation time $\tau_{\mathrm{in}}$.
 Since $\tau_{\mathrm{in}} \gg \tau_{\mathrm{el}}$ it may significantly exceed the conventional contribution.
This contribution to the linear conductivity exists only in the presence of a \emph{dc}  supercurrent. Furthermore, this contribution  is strongly anisotropic and depends on the relative orientation between  $\bm{E}_\omega$  and the supercurrent. Even in situations where this contribution is small in comparison to the conventional result, it determines  the dependence of the conductivity on both the magnitude and direction of the \emph{dc} supercurrent. This enables determination of $\tau_\mathrm{in}$ from  microwave absorption measurements.

The physical mechanism of this contribution to the conductivity is similar to the Debye mechanism of microwave absorption in gases \cite{debye_polar_1970}, Mandelstam-Leontovich mechanism  of the second viscosity in liquids \cite{landau_fluid_2013}, and Pollak-Geballe mechanism of microwave absorption in the hopping conductivity regime \cite{pollak_low-frequency_1961}.    It arises from the motion of energy levels of the system in the presence of the external field. As a result  of this motion the system deviates from thermal equilibrium.  In this case the equilibration is caused by energy relaxation processes and  the corresponding contribution to the conductivity is proportional to the energy relaxation time.

The physical origin of this mechanism in superconductors
 can be qualitatively understood as follows. Let us separate the superfluid momentum $\bm{ p}_\mathrm{s}(t)=\bar{\bm{p}}_\mathrm{s}+\delta \bm { p}_\mathrm{s}(t)$ into a \emph{dc} part $\bar{\bm{p}}_\mathrm{s}$ and an \emph{ac} part $\delta \bm { p}_\mathrm{s}(t)$, whose time evolution is determined by the microwave field
\begin{equation}
  \label{eq:acceleration}
 \delta \dot{\bm{p}}_{s}(t) =e{\bf E} (t).
\end{equation}
At low frequencies, $\omega \ll \tau_{\mathrm{el}}^{-1}$, the quasiparticle distribution function $n(\epsilon,t)$ depends only on the energy $\epsilon$ and time, while the  density of states per unit energy, $\nu(\epsilon, p_\mathrm{s})$,  depends on the instantaneous value of the superfluid momentum $p_\mathrm{s}$. As the value of ${\bf p}_{\mathrm s}$ changes with time, individual quasiparticle levels move in energy space. At finite temperature the quasiparticles occupying these levels travel in energy space as well. This motion creates a non-equilibrium quasiparticle distribution, which relaxes due to inelastic scattering causing entropy production and energy dissipation.  The corresponding contribution to the conductivity is proportional to $\tau_{\mathrm{in}}$.  The reason why the Debye contribution to the linear conductivity exists only at $\bar{\bm{p}}_\mathrm{s} \neq 0$ is the following.  Being invariant under time reversal the density of states must be an even function of the condensate momentum, and thus can depend only on  $|\bm{p}_\mathrm{s}|^2$  in an isotropic system.
As a result, in the linear in $\bm{E}$ approximation $\nu(\epsilon)$ changes in time proportionally to $\delta \bm{p}_\mathrm{s}(t) \cdot \bar{\bm{p}}_\mathrm{s}$.

\section{Relation between
the Debye contribution to the conductivity
and  the ${\bf p}_{\mathrm s}$-dependence of the density of states}

 In this section we show that the Debye contribution to the conductivity can be expressed in terms of the quasiparticle density of states in the presence of a supercurrent.

Below we assume the condensate momentum $\bm{p}_{\mathrm s}$ to be spatially uniform. This situation is realized in superconducting films with thickness smaller than the penetration length of the magnetic field $\lambda_{\mathrm H}$.
 We discuss applicability of our results to the case of bulk samples in Sec.~\ref{sec:conclusion}.
We also assume that the frequency of the microwave radiation satisfies the condition $\omega \ll \Delta$, where $\Delta$ is the pairing gap in the superconductor. In this regime we can describe the quasiparticles by an instantaneous energy spectrum, which depends on the value of $p_{\mathrm s}(t)$.
To describe the time evolution of the instantaneous energy levels we note that the number of levels in the system is conserved.  Therefore the density of states $\nu(\epsilon, p_\mathrm{s}(t))$
is subject to the continuity equation in energy space
\begin{equation}\label{eq:levelContinuity}
\partial_t \nu(\epsilon, p_{\mathrm{s}}) +\frac{\partial [v_{\nu} (\epsilon, p_{\mathrm{s}})\nu (\epsilon, p_{\mathrm{s}})]}{\partial \epsilon} =0,
\end{equation}
where $v_\nu (\epsilon, p_{\mathrm s})$ is the level ``velocity'' in energy space.
Using the condensate acceleration equation \eqref{eq:acceleration} the latter  can be expressed in the form
\begin{equation}\label{eq:level_continuity}
v_{\nu} (\epsilon, p_{\mathrm{s}})=  e \bm{E} \cdot \bm{V}(\epsilon,\bm{p}_{\mathrm s})
\end{equation}
 where
\begin{equation}\label{eq:level_velocity}
   \bm{V}(\epsilon,\bm{p}_{\mathrm s}) = - \frac{1}{\nu (\epsilon,p_{\mathrm s})}  \int_{0}^{\epsilon} d \tilde{\epsilon}  \frac{\partial \nu (\tilde{\epsilon}, p_{\mathrm s} )}{\partial \bm{p}_{\mathrm s}}
\end{equation}
characterizes the sensitivity of the energy levels to changes of  ${\bf p}_{\mathrm{s}}$.

In the regime $\omega, \tau_{\mathrm{in}}^{-1}\ll \tau_{\mathrm{el}}^{-1}$  the  quasiparticle distribution function $n(\epsilon,t)$, which describes the occupancy of the quasiparticle energy levels,   depends only on the energy.  In the absence of inelastic scattering its time evolution due to the spectral flow is described by the continuity equation $\partial_t(\nu n)+\partial_\epsilon(v_{\nu} \nu n) =0$. Combining this with the continuity equation \eqref{eq:levelContinuity}
for $\nu (\epsilon, p_s)$ and allowing for inelastic collisions
we obtain the kinetic equation
\begin{equation}\label{eq:n_dot}
 \partial_{t} n (\epsilon, t)+ e\bm{E}(t)\cdot  \bm{V} (\epsilon, \bm{p}_{\mathrm s})\,  \partial_\epsilon  n(\epsilon, t) = I_{\mathrm{in}}\{  n\},
\end{equation}
where $I_{\mathrm{in}}\{ n\}$ is the collision integral describing inelastic  scattering of quasiparticles.

The power $W$ of microwave radiation absorbed per unit volume of the superconductor may be obtained by evaluating the rate of work performed by the microwave field on the quasiparticles, which is given by
\begin{equation}\label{eq:absorption power}
  W = \int_0^\infty d \epsilon \big\langle \nu (\epsilon, p_{\mathrm{s}} (t))n(\epsilon, t)   e\bm{E}(t)\cdot  \bm{V} (\epsilon, \bm{p}_{\mathrm s}(t))\big\rangle.
\end{equation}
Here $\langle \ldots \rangle$ denotes time averaging.  Below we characterize the absorption power by the dissipative part of the conductivity $\sigma_\mathrm{DB} $ defined by
\begin{equation}\label{eq:sigma_W}
  \frac{\sigma_\mathrm{DB}}{2} \,  E_\omega^2 = W.
\end{equation}

In the relaxation time approximation the scattering integral describing the inelastic quasiparticle scattering can be written as
\begin{equation}\label{eq:I_relaxation_time}
  I_{\mathrm{in}}\{ n\}= -\frac{\delta n(\epsilon,t)}{\tau_{\mathrm{in}}},
\end{equation}
where $\delta n (\epsilon) = n (\epsilon ) - n_\mathrm{F}(\epsilon)$,  with $n_{\mathrm{F}} (\epsilon)$  being the Fermi function, is the nonequilibrium part of the quasiparticle distribution.

Using the condensate acceleration equation \eqref{eq:acceleration} we obtain in the Fourier representation
\[
\delta n (\epsilon) = - \frac{e \bm{E}\cdot \bm{V} (\epsilon, \bar{\bm{p}}_{\mathrm{s}})}{ - i \omega + \tau_{\mathrm{\mathrm{in}}}^{-1} } \frac{d n_{\mathrm F}(\epsilon)}{d \epsilon}.
\]
Substituting this expression into  Eq.~\eqref{eq:absorption power} we obtain the following expression for the  real part of Debye contribution to the \emph{ac} conductivity

\begin{equation}
\label{eq:sigma_ratio}
\frac{\sigma_{\mathrm{DB}}}{\sigma_\mathrm{D}}=
 \frac{3\tau_{\mathrm{in}}}{4\tau_{\mathrm{el}}} \frac{1}{\left[ 1 + \left( \omega \tau_{\mathrm{in}}\right)^2\right]}    \int_0^\infty   \frac{ d \epsilon}{T} \frac{ \nu (\epsilon,\bar{p}_{\mathrm s})
 V^2(\epsilon,\bar{p}_{\mathrm s})}{ \nu_n v_{\mathrm F}^2 \cosh^{2}(\epsilon/2T)}.
\end{equation}

Equation \eqref{eq:sigma_ratio} expresses the Debye contribution to the conductivity in terms of the density of states in a current-carrying superconductor.  Both the kinetic scheme based on  Eqs.~\eqref{eq:level_velocity}, \eqref{eq:n_dot} and \eqref{eq:absorption power}, and Eq.~\eqref{eq:sigma_ratio} for the Debye contribution to the conductivity are general: they apply to superconductors with arbitrary symmetry of the order parameter.  They also account for broadening of the mean field features of the density of states, which could be  due to non-uniformity of the interaction constant, inelastic scattering and quantum and classical fluctuations of the order parameter.

It follows from Eq.~\eqref{eq:sigma_ratio} that energy relaxation processes affect the electric conductivity. This happens because in the presence of \emph{dc}-supercurrent the energy of a quasiparticle state depends on the current carried by it.  To elucidate this issue in Sec.~\ref{sec:clean} we rederive Eq.~\eqref{eq:sigma_ratio} by obtaining an explicit expression for the electric current. We focus on a particular case of clean superconductors, where the elastic mean free path is larger than the superconducting coherence length.

\subsection{Derivation of Eq.~\eqref{eq:sigma_ratio} for clean superconductors} \label{sec:clean}

In clean superconductors where the mean free path exceeds the superconducting coherence length the nonequilibrium state of the superconductor  may be described by  the quasiparticle distribution function $n_{\bm{p}}$.
In this case the current density is expressed in terms of the quasiparticle distribution function as
\begin{equation}\label{eq:current_n_p}
 \bm{j} = e N\frac{\bm{p}_{\mathrm s}}{m} + 2 e\int \frac{d^3 p}{(2\pi)^3}
 \bm{v} \, n _{\bm{p}}.
\end{equation}
Here $N$ is the electron density and $\bm{v} = \bm{p}/m$ is the band velocity of the electron  with quasimomentum $\bm{p}$.

The time evolution of the distribution function is described by the Boltzmann  kinetic equation, which in the spatially uniform case takes a simple form
\begin{eqnarray}\label{eq:kineq}
   \partial_t  n _{\bm{p}} =  I_{\mathrm{el}} + I_{\mathrm{in}}.
\end{eqnarray}
Here $I_{\mathrm{el}}$ and $I_{\mathrm{in}}$ are the collision integrals describing, correspondingly, the elastic and inelastic scattering processes.

     The reason the conductivity is affected by the inelastic collisions is that in the presence of supercurrent the quasiparticle energy spectrum,
\begin{equation}\label{eq:quasiparticle_energy}
    \tilde{\epsilon}_{\bm{p}} (\bm{p}_{\mathrm s}) =\sqrt{|\Delta(\bm{p})|^2 + \xi^2_{\bm{p}}} + \bm{p}_{\mathrm s}   \cdot \bm{v},
\end{equation}
contains an odd-in-momentum part  described by the second term above.
 Since we are interested in the regime $\tau_{\mathrm{el}}\ll \tau_{\mathrm{in}}$ ,
 $\omega \tau_{\mathrm{el}} \ll 1$ the quasiparticle distribution function depends only on the quasiparticle energy  $ n_{\bm{p}}= n\left(\tilde{\epsilon}_{\bm{p}} (\bm{p}_{\mathrm s} ), t\right)$.  Substituting  this form into Eq.~\eqref{eq:current_n_p}, noting that
$\bm{v} = \frac{d}{d \bm{p}_{\mathrm s} }    \tilde{\epsilon }_{\bm{p}} (\bm{p}_{\mathrm s} )$,  and using the resolution of identity $1= \int_0^\infty d\epsilon \delta[\epsilon - \tilde{\epsilon}_{\bm{p} } (\bm{p}_{\mathrm s} )]$  we can express the current density as
\begin{equation}\label{eq:current_n_epsilon}
 \bm{j} = e N\frac{\bm{p}_{\mathrm s}}{m} +  e\int_0^\infty d\epsilon \,
  n (\epsilon, t)     \nu(\epsilon, \bm{p}_{\mathrm s} ) \bm{V} (\epsilon, \bm{p}_{\mathrm s} ),
\end{equation}
where
\begin{equation}\label{eq:nu_def}
  \nu(\epsilon, \bm{p}_{\mathrm s} )= 2\int \frac{d^3 p}{(2\pi \hbar)^3} \delta[\epsilon - \tilde{\epsilon}_{\bm{p}} (\bm{p}_{\mathrm s} )].
\end{equation}
is the density of states, and

\begin{equation}\label{eq:V_epsilon_clean}
  \bm{V} (\epsilon, \bm{p}_{\mathrm s} ) = \frac{1}{\nu[\epsilon, \bm{p}_{\mathrm s} ]}        \int \frac{d^3 p}{(2\pi \hbar)^3}   \delta [ \epsilon - \tilde{\epsilon}_{\bm{p}} (\bm{p}_{\mathrm s} )] \,      \frac{d}{d \bm{p}_{\mathrm s} }    \tilde{\epsilon }_{\bm{p}} [\bm{p}_{\mathrm s} ].
\end{equation}
 Writing the $\delta$-function in the integrand above as a derivative of the step-function, and integrating by parts it is easy to show that Eq.~\eqref{eq:V_epsilon_clean} reduces to Eq.~\eqref{eq:level_velocity}.   Thus Eq.~\eqref{eq:current_n_epsilon} expresses the current density in terms of the energy-dependent distribution function $n (\epsilon, t) $ and $p_{\mathrm s}$-dependence of the density of states.

Finally, in order to obtain the time evolution equation for $n (\epsilon, t) $
 we substitute the distribution function in the form  $ n_{\bm{p}}= n\left(\tilde{\epsilon}_{\bm{p}} (\bm{p}_{\mathrm s}), t\right)$  into Eq.~\eqref{eq:kineq}, multiply it by $\delta [\epsilon - \tilde{\epsilon}_{\bm{p}} (\bm{p}_{\mathrm s})]$  and integrate over $\frac{d^3 p}{(2\pi)^3}$.
Then, using the fact that $\partial_t n _{\bm{p}} = \partial_t n\left(\tilde{\epsilon}_{\bm{p}} , t\right) +  \bm{v}\cdot \dot{\bm{p}}_{\mathrm s}    \, \partial _{\tilde{\epsilon}_{\bm{p}} }   n\left(\tilde{\epsilon}_{\bm{p}} , t\right) $, and noting that the elastic collision integral is nullified by an arbitrary distribution function that depends only on  $\tilde{\epsilon}_{\bm{p}} (\bm{p}_{\mathrm s})$ we reproduce Eq.~\eqref{eq:n_dot}. Linearizing it and substituting the result for $\delta n$ into Eq.~\eqref{eq:current_n_p} we get Eq.~\eqref{eq:sigma_ratio}.

\section{Microwave conductivity in $s$- and $d$-wave superconductors \label{sec:s_and_d}}

In this section we use Eq.~\eqref{eq:sigma_ratio} to obtain expressions for the Debye contribution to the conductivity of $s$- and $d$- superconductors. We focus on the case of small values of the supercurrent. We will show that in this limit  the $p_{s}$-dependence of $\sigma_{\mathrm{DB}}(p_{\mathrm s})$ is stronger than quadratic, that is $\sigma_{\mathrm{DB}}(p_{\mathrm s})/p_{\mathrm s}^2  \to \infty$ at $p_{\mathrm s} \to 0$. Therefore we neglect the $p_{s}$-dependence of the order parameter, as its contribution to $\sigma_{\mathrm{DB}}$ is quadratic in $p_{s}$.
 In  Sec.~\ref{sec:T_c} we start with the  regime of temperatures close to the critical temperature, $|T-T_{\mathrm c}|\ll T_{\mathrm c}$. In Sec.~\ref{sec:low_T} we consider the low temperature regime, $T\ll T_\mathrm{c}$.

\subsection{Regime of temperatures near the critical temperature \label{sec:T_c}}

 At  $|T-T_\mathrm{c}|\ll T_\mathrm{c}$
the density of states is affected by the condensate momentum
in a narrow energy window $|\epsilon -\Delta|\ll T$.  Since the energy transfer in a typical inelastic collision is of order $T$ the relaxation time approximation for the inelastic collision integral Eq.~\eqref{eq:I_relaxation_time} is asymptotically exact, while the relaxation time $\tau_{\mathrm{in}}(T)$  depends only on the temperature $T$.

\subsubsection{ $s$-wave superconductors}

We start with a discussion of the  Debye contribution to the \emph{ac} conductivity
of $s$-wave superconductors, see Ref.~\cite{smith_debye_2019}.  For an isotropic spectrum, which we assume below, the vector $\bm{V} (\epsilon,\bm{p}_{\mathrm s})$  in Eq.~\eqref{eq:level_velocity} is parallel to $\bm{p}_{\mathrm s}$.
In this case only the longitudinal conductivity,  which corresponds to $\bm{E}_\omega \parallel \bar{\bm{p}}_{\mathrm s}$, is affected by inelastic relaxation.

The density of states is most strongly affected by the supercurrent at energies near the  gap $\Delta$. Namely at $\bar{p}_\mathrm{s}\neq 0$ the peak in the BCS density of states, $\nu (\epsilon, 0) \to \nu_n \sqrt{\frac{\Delta}{2(\epsilon - \Delta)}}$  at  $\epsilon \to \Delta$, is broadened. The width of the broadening and the shape of the peak depend on the magnitude of the condensate momentum $\bar{p}_{\mathrm s}$ and the strength of disorder.

\textit{Ballistic regime.---} In the regime $v_{\mathrm F} \bar{p}_{\mathrm s}\tau_{\mathrm{el}}^2\Delta \gg 1 $, (which can be realized only in clean superconductors, $\Delta \tau_\mathrm{el} \gg 1$) the density of states, $\nu(\epsilon ,  p_{s})$, can be found using the standard expression Eq.~\eqref{eq:quasiparticle_energy} for the quasiparticle spectrum.
  In the relevant energy interval $|\epsilon - \Delta | \ll \Delta$ one obtains
\begin{equation}\label{eq:purenu}
  \frac{\nu(\epsilon ,  p_{s})}{\nu_n} = \sqrt{\frac{\Delta}{2v_{\mathrm F} p_{\mathrm s}}}\left[\theta(z+1)\sqrt{z+1}  - \theta(z-1)\sqrt{z-1} \right],
\end{equation}
where $z = (\epsilon - \Delta)/v_{\mathrm F} p_\mathrm{s} $, and  $\theta (z)$ is the Heavyside step-function.
The width of the broadening of the BCS peak is $\delta \epsilon \sim v_{\mathrm F} \bar{p}_\mathrm{s}$.  Using Eq.~\eqref{eq:level_velocity}  and  Eq.~\eqref{eq:sigma_ratio}
we obtain for the Debye contribution to the conductivity in the ballistic regime
 \begin{equation}\label{eq:sigma_ballistic}
  \frac{\sigma_\mathrm{DB}}{\sigma_{\mathrm{D}}} =  I_{\mathrm b} \frac{\tau_{\mathrm{in}}}{\tau_{\mathrm{el}}  \left[ 1 + (\omega \tau_\mathrm{in})^2 \right]} \frac{\Delta}{T}  \sqrt{\frac{v_\mathrm{F}  \bar{p}_\mathrm{s} }{\Delta}}
\end{equation}
where $I_{\mathrm b}=\frac{8}{45}$.

Eq.~\eqref{eq:purenu} for the density of states is valid  as long as the broadening due to elastic scattering $\tau^{-1}_\mathrm{el} (\epsilon)$ is smaller than the relevant energy interval in the problem $|\epsilon - \Delta | \lesssim v_\mathrm{F} \bar{p}_\mathrm{s}\ll \Delta $.
  Here $\tau_\mathrm{el} (\epsilon)$ is the energy-dependent quasiparticle  mean free time, which for $|\epsilon  -\Delta| \ll \Delta$ is given by the standard expression
 \begin{equation}
 \tau^{-1}_\mathrm{el} (\epsilon) \approx \tau^{-1}_\mathrm{el}\sqrt{\frac{2(\epsilon - \Delta)}{\Delta}}
 \end{equation}
 (see for example \cite{mineev_introduction_1999}). Therefore the regime of ballistic motion of quasiparticles participating in the Debye mechanism of microwave absorption is realized at relatively  large supercurrent densities, where
 \begin{equation}
 v_\mathrm{F} \bar{p}_\mathrm{s}\tau_{\mathrm{el}}^2\Delta \gg 1 .
 \end{equation}

\emph{Diffusive regime.---} In the opposite  limit  $v_\mathrm{F} \bar{p}_\mathrm{s}\tau_{\mathrm{el}}^2\Delta \ll 1$   the quasiparticles participating in the Debye absorprtion mechanism move diffusively, and disorder may no longer be ignored.\footnote{It is worth noting that the diffusive regime can be realized in both clean, $\Delta \tau_\mathrm{el} \ll 1$, and dirty $\Delta \tau_\mathrm{el} \gg 1$, superconductors.} In this case the Debye contribution to the conductivity  can be studied using the standard theoretical methods developed in the theory of disordered superconductors~\cite{ parks_superconductivity:_1969,abrikosov_methods_1975,larkin_nonlinear_1977}.
The quasiparticle density  of states can be written as
\begin{equation}\label{eq:nuDisord}
    \nu(\epsilon) = \nu_n \Re \left\langle \frac{\bar{\epsilon} + \bm{v} \cdot \bm{p}_{\mathrm s}}{\sqrt{(\bar{\epsilon} + \bm{v}\cdot \bm{p}_{\mathrm s})^2 - |\bar{\Delta}(\bf{k})|^2}}\right\rangle
    \end{equation}
where $\langle \ldots \rangle$ denotes averaging over the Fermi surface, and $\bar{\epsilon}$ and $\bar{\Delta}({\bf k})$ are the disorder-renormalized energy and order parameter respectively.
For example, in the case of a white noise disorder in the Born approximation they are given by~\cite{abrikosov_methods_1975,parks_superconductivity:_1969}
\begin{subequations}\label{eq:g_def}
\begin{eqnarray}
 \bar{\epsilon} & = & \epsilon + \frac{i}{2\tau_{\mathrm{el}} }  \left\langle \frac{\bar{\epsilon} + \bm{v}\cdot \bm{p}_{\mathrm s}}{\sqrt{( \bar{\epsilon}+ \bm{v}\cdot \bm{p}_{\mathrm s})^2 - |\bar{\Delta}({\bf k}) |^2 }} \right\rangle  \\
 \bar{\Delta}({\bf k}) & = &  \Delta ({\bf k}) + \frac{i}{2\tau_{\mathrm{el}}}
  \left\langle \frac{ \bar{\Delta}({\bf k})}{\sqrt{( \bar{\epsilon}+ \bm{v}\cdot \bm{p}_{\mathrm s})^2 - |\bar{\Delta}({\bf k}) |^2 }}\right\rangle.
\end{eqnarray}
\end{subequations}
We have shown in Ref.~\cite{smith_debye_2019} that for isotropic $s$-wave superconductors the density of states can be expressed as
\begin{equation}
\label{eq:IntermediateDOS}
    \frac{\nu(\epsilon,\bar{p}_{\mathrm s})}{\nu_n} = \frac{1}{\sqrt{2}}\Im y^{-1},
\end{equation}
 where $y$ is determined by the following equation
\begin{equation}
\label{eq:Cubic Eqn}
    y\left(y^2+w\right) +\frac{\sqrt{2}\zeta^2}{3 \gamma} = 0.
\end{equation}
Here $\zeta = v_{\mathrm{F}} \bar{p}_{\mathrm{s}}/\Delta$, $\gamma = (\tau_{\mathrm{el}} \Delta)^{-1}$, and $w=(\epsilon-\Delta)/\Delta$. The solutions of this equation can be written in the scaling form $y= \frac{\zeta^{2/3}}{\gamma^{1/3}} \, \tilde{y} \left( \frac{w \gamma^{2/3}}{\zeta^{4/3}} \right) $.
  Therefore in this case the width of the broadening of the BCS peak is
 $\delta \epsilon (\bar{p}_\mathrm{s}) \sim \left(\Delta D^2 \bar{p}_\mathrm{s}^4\right)^{1/3}$.
 The explicit form of  $\tilde{y} \left( \frac{w \gamma^{2/3}}{\zeta^{4/3}} \right)$ is given by the Cardano formula, (See Eq. (S.17) in Ref.~\cite{smith_debye_2019}).
Substituting this form
into Eq.~\eqref{eq:IntermediateDOS}, and using Eqs.~\eqref{eq:level_velocity} and \eqref{eq:sigma_ratio}, we obtain
\begin{equation}\label{eq:sigma_diffusive}
  \frac{\sigma_\mathrm{DB} }{\sigma_{\mathrm{D}}} = I_{\mathrm d} \frac{\tau_{\mathrm{in}}}{\tau_\mathrm{el}} \frac{\Delta}{T} \frac{\tau_\mathrm{el}\left(\Delta D^2 \bar{p}_{\mathrm s}^4 \right)^{1/3}}{\left[1 + \left( \omega \tau_{\mathrm{in}}\right)^2\right]},
\end{equation}
where $I_{\mathrm d} \approx  0.0549$. This expression is consistent with the result obtained in Ref.~\cite{ovchinnikov_electromagnetic_1978} by a different method.

The nonanalytic dependences of $\sigma_\mathrm{DB}$ on $\bar{p}_\mathrm{s}$ in Eqs.~\eqref{eq:sigma_ballistic} and
\eqref{eq:sigma_diffusive}
are related to the divergence of the BCS density of states at $\epsilon = \Delta$.
In real superconductors this divergence is smeared by  pairbreaking processes and non-uniformity of the electron interaction constant, which are characterized by a broadening energy  scale $\Gamma \ll | \Delta |$.  Consequently,  at $\delta \epsilon (\bar{p}_\mathrm{s})\ll \Gamma$ the $\bar{p}_\mathrm{s}$-dependence of the conductivity should become analytic, $\sigma_{DB} = c \,  \bar{p}_\mathrm{s}^2$.
The magnitude of the coefficient $c$ can be estimated by matching this expression to Eqs.~\eqref{eq:sigma_ballistic} and \eqref{eq:sigma_diffusive} at the values of $\bar{p}_\mathrm{s}$  determined by
 the condition that  the  energy broadening of the BCS singularity, $\delta \epsilon (\bar{p}_\mathrm{s})$ be of order   $\Gamma$. This yields
\begin{equation}
  \frac{\sigma_{\mathrm{DB}}}  {\sigma_{\mathrm D}}\sim \left(\frac{v_{\mathrm F}\bar{p}_{\mathrm s}}{\Gamma}\right)^{2}\frac{\Delta}{T} \frac{\tau_{\mathrm{in}}}{\tau_{\mathrm{el}}  \left[ 1 + (\omega \tau_\mathrm{in})^2 \right]}
    \begin{cases}
    \sqrt{\frac{\Gamma }{\Delta}} & \text{for} \,\
v_\mathrm{F} \bar{p}_\mathrm{s}\tau_{\mathrm{el}}^2\Delta \gg 1 ,\\
  \tau_\mathrm{el}^2\sqrt{\Delta \Gamma^{3}  } &
   \text{for} \,\  v_\mathrm{F} \bar{p}_\mathrm{s}\tau_{\mathrm{el}}^2\Delta \ll 1   . \end{cases}
\end{equation}

\subsubsection{$d$-wave superconductors \label{sec:T_c_d-wave}}

Let us now apply the general expression \eqref{eq:sigma_ratio} to study the Debye contribution to the conductivity of $d$-wave superconductors.  {The order parameter in $d$-wave superconductors $\Delta({\bf p})$ changes its sign upon rotation of the momentum by $\pi/2$ in the $xy$ plane, and can be modeled by the form
\begin{equation}
\Delta(\bm{p})=\Delta_{0}(\sin^{2} p_{x}a -\cos^{2} p_{y}a),
\end{equation}
where $\Delta_{0}(T,\tau_{\mathrm{el}})$ is the gap maximum at the antinode, which generally depends on temperature and $\tau_{\mathrm{el}}$.  In this article we focus on the limit $\Delta_{0}\tau_{\mathrm{el}}\gg 1$.
In this case the density of states in the presence of supercurrent may be evaluated with the aid of  Eqs.~\eqref{eq:nuDisord} and \eqref{eq:g_def}.
The integral in Eq.~\eqref{eq:sigma_ratio} for the Debye contribution to the conductivity is dominated by a narrow energy interval $|\epsilon - \Delta_0|\ll \Delta_0$, which corresponds to quasiparticles with momenta near the antinodes.

Let us begin with the clean limit, $\tau_{\mathrm{el}}\rightarrow \infty$. In this case the density of states may be evaluated using Eq.~\eqref{eq:nu_def}.
For $|\epsilon - \Delta_0|\ll \Delta_0$ we obtain
\begin{equation}\label{eq:nu_d_anti}
 \nu(\epsilon, p_{\mathrm{s}}) = \frac{\nu_n}{\pi} \sum_i
    \ln \frac{\Delta_{0}}{|(\epsilon-\Delta_{0})+v_{\mathrm F}(\bm{n}_i\cdot\bm{p}_{\mathrm{s}})|},
\end{equation}
where  the summation is performed over all antinodal lines and $\bm{n}_i$  is the unit vector in the direction of the $i$-th antinodal line.

{The energy level sensitivity $\bm{V} (\epsilon)$ in the clean limit may be determined from Eq.~\eqref{eq:V_epsilon_clean}, and is given by
\begin{equation}\label{eq:DVpureAntinode}
\bm{V}(\epsilon)
  = v_{\mathrm F} \frac{\sum_i (\bm{n}_i\cdot \hat{\bm{p}}_{\mathrm{s}})\bm{n}_i \ln\left( \frac{\Delta_0}{|(\epsilon - \Delta_0) + v_{\mathrm F} (\bm{n}_i \cdot \bm{p}_{\mathrm s})|}\right)}{\sum_i
    \ln \frac{\Delta_{0}}{|(\epsilon-\Delta_{0})+v_{\mathrm F}(\bm{n}_i\cdot\bm{p}_{\mathrm s})|}}   .
\end{equation}
Substituting Eqs.~\eqref{eq:nu_d_anti},  and \eqref{eq:DVpureAntinode}  into Eq.~\eqref{eq:sigma_ratio} and assuming $T \gg v_{\mathrm F} \bar{p}_{\mathrm s}$ within logarithmic accuracy we obtain the following expression for the Debye contribution to the conductivity,
\begin{equation}\label{eq:pureSigmaDTc}
     \frac{\sigma_{\mathrm{DB}}}{\sigma_{\mathrm D}} = \frac{3}{\pi} \frac{\tau_{\mathrm{in}}}{\tau_{\mathrm{el}}}\frac{1}{\left[1+(\omega\tau_{\mathrm{in}})^2\right]}\left(\frac{v_{\mathrm F} \bar{p}_{\mathrm s}}{T} \right) \ln\left(\frac{\Delta_{0}}{v_{\mathrm F} \bar{p}_{\mathrm s}} \right).
     \end{equation}
To derive this result we neglected the contributions of quasiparticles near the nodal lines to $\sigma_{\mathrm{DB}}$  because they are small  in the ratio $v_F \bar{p}_{\mathrm s}/\Delta$ as compared to that in Eq.~\eqref{eq:pureSigmaDTc}.

Equations \eqref{eq:nu_d_anti} and \eqref{eq:pureSigmaDTc} are valid provided $v_{\mathrm F}\bar{p}_{\mathrm s} > \tau_{\mathrm{el}}^{-1}$.
In the presence of disorder the non-analyticity of the density of states  as a function of $\epsilon$, Eq.~\eqref{eq:nu_d_anti}, is smeared in the interval of energies of order $\tau_{\mathrm{el}}^{-1}$.
In the limit of small supercurrent,   $v_{\mathrm F}\bar{p}_{\mathrm s}\ll \tau_{\mathrm{el}}^{-1}$,  the Debye contribution to the conductivity is expected to be analytic in $\bar{p}_s$, namely $\sigma_{\mathrm{DB}}\sim a \bar{p}_{\mathrm s}^{2}$. The value of the coefficient $a$  can be estimated by matching this expression with Eq.~\eqref{eq:pureSigmaDTc}  at $v_{\mathrm F}\bar{p}_{\mathrm s} \sim \tau_{\mathrm{el}}^{-1}$. This yields
\begin{equation}
     \frac{\sigma_{\mathrm{DB}}}{\sigma_{\mathrm D}} \sim \frac{\tau_{\mathrm{in}}}{T}\frac{(v_{\mathrm F} \bar{p}_{\mathrm s})^2}{\left[1+(\omega\tau_{\mathrm{in}})^2\right]} .
\end{equation}
In the Born approximation this result can be obtained from Eqs.~\eqref{eq:nuDisord} and ~\eqref{eq:g_def}.

\subsection{Low temperature regime, $T\ll \Delta_{0}$ \label{sec:low_T} }

Low temperature quasiparticle kinetics in $s$- and $d$-wave superconductors have common features.  In both cases the low energy density of states is suppressed. Therefore, in both cases the quasiparticle concentration decreases with temperature more rapidly than in normal metals. Consequently the electron-electron scattering rate is suppressed and the quasiparticle energy relaxation is controlled by electron-phonon scattering.

Furthermore,  one needs to  distinguish between two different types of  inelastic scattering processes in superconductors.  The quasiparticle-phonon relaxation processes that conserve the number of quasiparticles are characterized by the rate $1/\tau_{\mathrm{in}}^{(st)}(T)$,  which is independent of quasiparticle concentration.\footnote{ We note that  in $d$-wave superconductors  the value of $\tau_{\mathrm{in}}^{(st)}(T)\sim \Theta_{\mathrm D}^{2}/T^{3}$  is of the same order as  that in normal metals} The second type of inelastic relaxation processes corresponds to  recombination, which changes the total number of quasiparticles.
The rate $ 1/\tau_{\mathrm r} (T)$ of such processes is proportional to the quasiparticle concentration $x(T)$.  Therefore at low temperatures it becomes much smaller than  $1/\tau_{\mathrm{in}}^{(st)}(T)$;
\begin{equation}\label{eq:tau_recombination}
 \tau_{\mathrm r} (T) \propto \frac{\tau_{\mathrm r}^{(0)} (T)}{x(T)}\gg \tau_{\mathrm{in}}^{(st)} (T).
\end{equation}

The Debye contribution to the dissipative kinetic coefficients  is proportional to the longest relaxation time in a system (see for example \cite{landau_fluid_2013}), which  in our case  is
$\tau_{\mathrm r} (T)$.   On the other hand $\sigma_{\mathrm{DB}}$ is also proportional to the density of thermal quasiparticles. We show below that, as a consequence,  the Debye contribution to the conductivity becomes independent of the quasiparticle concentration $x (T)$. As a result,  its magnitude  in the low temperature  regime is roughly speaking of the same order as  that near $T_{\mathrm c}$.

In order to obtain an estimate for $\sigma_{\mathrm{DB}}$ in this regime we note that
since the recombination time is the longest time scale in the problem, $\tau_{\mathrm r}\gg \tau_{\mathrm{in}}^{(st)}$, at relatively short time scales of order of $\tau_{\mathrm{in}}^{(st)}$
the number of quasiparticles is approximately conserved. As a result, at such time scales the system of quasiparticles reaches a  quasi-equilibrium form which is characterized by a nonzero chemical potential,
\begin{equation}
n(\epsilon)=   \frac{1}{1+\exp(\frac{\epsilon-\mu}{T})},
\end{equation}
while in thermal equilibrium $\mu=0$.  To find the value of $\mu$ in the presence of microwave radiation one has
to integrate Eq.~\eqref{eq:n_dot} over $\epsilon$ bearing in mind that the relaxation processes conserve the number of quasiparticles, $\int  I_{\mathrm{st}} d \epsilon =0$.
Doing so, we get the following estimate for the chemical potential
\begin{equation}\label{eq:mu}
\mu\sim \frac{\tau_{\mathrm r}}{n_{\mathrm F}(\epsilon^{*})}\int    e\bm{E}(t) \cdot  \bm{V} (\epsilon, \bm{p}_{\mathrm s})\,  \frac{ d n_{\mathrm F}(\epsilon)}{d \epsilon}   d \epsilon.
\end{equation}
Here $\epsilon^{*}=\Delta$ in the case of s-wave superconductors, and $\epsilon^{*}=0$ for the case of d-wave superconductors.
To get $\sigma_{\mathrm{DB}}$  one should substitute $\delta n (\epsilon)\sim \mu d n_{\mathrm F}(\epsilon)/d\epsilon$ into Eqs.~\eqref{eq:absorption power} and \eqref{eq:sigma_W}.
Since in this regime  the relaxation time approximation for the recombination collision integral is only applicable to accuracy within a factor of order unity,  both  Eq.~\eqref{eq:mu} and subsequent estimates for $\sigma_{\mathrm{DB}}$ are valid only with the same accuracy.

\subsubsection{$s$-wave superconductors \label{sec:s-wave_low}}

In \emph{s}-wave superconductors the dimensionless quasiparticle concentration $x_{\mathrm s}(T)$  defined by
\begin{equation}
 x_{\mathrm s}(T)=(\nu_n \Delta)^{-1}\int_{0}^{\infty} d \epsilon \nu (\epsilon ) n_{\mathrm F}(\epsilon) \sim \sqrt{ \frac{T}{\Delta}} \exp(-\Delta/T)
\end{equation}
is exponentially small. Consequently, the conventional contribution to the microwave absorption coefficient  is exponentially small as well.
On the other hand, since the recombination rate in Eq.~\eqref{eq:tau_recombination} is inversely proportional to the quasiparticle concentration,\footnote{ The parameter $\tau_{\mathrm r}^{(0)}$  in Eq.~\eqref{eq:tau_recombination}  may be estimated as  $1/\tau_{\mathrm r}^{(0)}\sim \Delta^{3}/\theta_{\mathrm D}^{2}$, where $\Theta_{\mathrm D}$ is  the Debye temperature.}  in the low frequency limit, $\omega\tau_{\mathrm r}\ll 1$, the exponentially small factor $\exp(-\Delta/T)$ is canceled from the expression for the conductivity.
Below we illustrate this fact in the diffusive regime, and  at $T \ll \delta \epsilon(\bar{p}_{\mathrm s}) \ll \Delta$.  In this case the magnitude of  the level sensitivity in the energy interval  $|\epsilon - \Delta| \lesssim T$  is $ V \sim \frac{1}{\bar{p}_\mathrm{s}}  \delta \epsilon \sim   \left( \Delta D^2 \bar{p}_\mathrm{s} \right)^{1/3} $. Thus, we get
\begin{equation}\label{eq:condDlowT}
\frac{\sigma_\mathrm{DB}}{\sigma_\mathrm{D}}\sim  \frac{\tau_{\mathrm r}^{(0)}}{\tau_{\mathrm{el}}}\sqrt{\frac{\Delta}{T}}\tau_{\mathrm{el}}\left(\Delta D^2 \bar{p}_{\mathrm s}^4\right)^{1/3}.
\end{equation}
We  note that the value of the conductivity at zero superfluid momentum may be estimated as $\sigma(\bar{p}_{s}=0)\sim x_{\mathrm s} (T)\sigma_{\mathrm D}$, and is exponentially small at $T\ll \Delta$.  Thus, in this regime the Debye contribution to the conductivity becomes exponentially enhanced at low temperatures  in comparison to the conventional contribution.

\subsubsection{$d$-wave superconductors \label{sec:d-wave_low}}

The low energy density of states in $d$-wave superconductors is dominated by momenta in the vicinity of the nodal  lines, and in the clean limit $\tau_{\mathrm{el}}\rightarrow \infty$ is given by~\cite{volovik_fermionic_1997}
\begin{equation}\label{eq:DnuNode}
    \nu(\epsilon, p_{\mathrm s}) = \nu_n \sum_{i} \frac{|\epsilon + v_{\mathrm F} (\bm{m}_i \cdot \bm{p}_{\mathrm s}) |}{\Delta_0},
\end{equation}
where $\bm{m}_i$ denotes the unit vector pointing in the direction of the $i$-th nodal line.
Using Eq.~\eqref{eq:level_velocity} we find that at $\epsilon \ll \Delta_0$  the level sensitivity
$\bm{V} (\epsilon)$ is given by
\begin{equation}\label{eq:DVpureNode}
    \bm{V} (\epsilon)
  = v_{\mathrm F} \, \frac{\sum_i \bm{m}_i (\bm{m}_i \cdot \hat{\bm{p}}_{\mathrm s}) |\epsilon + v_{\mathrm F} (\bm{m}_i \cdot \bm{p}_{\mathrm s})|}{\sum_i |\epsilon + v_{\mathrm F} (\bm{m}_i \cdot \bm{p}_{\mathrm s})| }.
\end{equation}
Substituting Eqs.~\eqref{eq:DnuNode} and~\eqref{eq:DVpureNode} into Eq.~\eqref{eq:sigma_ratio} we find
\begin{equation}\label{eq:sigmaBDsmallT}
  \frac{\sigma_{\mathrm{DB}}}{\sigma_{\mathrm D}}\sim \frac{\tau_{\mathrm r } (T)}{\tau_{\mathrm{el}}}\frac{1}{\left[1+\left(\omega \tau_{\mathrm r}(T)\right)^{2}\right]}\begin{cases}
 \left(\frac{v_{\mathrm F} \bar{p}_{\mathrm s}}{\Delta_0} \right)^2 \frac{\Delta_0}{T}   \ln\left(\frac{T}{v_{\mathrm F} \bar{p}_{\mathrm s}} \right) & \mbox{ for } T\gg v_{\mathrm F}\bar{p}_{\mathrm s}, \\
\frac{T^{2}}{v_{\mathrm F}\bar{p}_{\mathrm s}\Delta_{0}}
 &  \mbox{ for }  T  \ll  v_{\mathrm F}\bar{p}_{\mathrm s} . \end{cases}
\end{equation}
The  recombination time here may be estimated using Eq.~\eqref{eq:tau_recombination} by noting that in \emph{d}-wave superconductors the dimensionless quasiparticle concentration decreases only as a power law in $T$
 \begin{equation}\label{eq:x_d_clean}
 x_{\mathrm d}(T)=(\nu_n T)^{-1}\int_{0}^{\infty} d \epsilon \nu (\epsilon ) n_{\mathrm F}(\epsilon) \sim \frac{T}{\Delta_{0}},
 \end{equation}
while  $\tau_{\mathrm{r}}^{(0)}$ in Eq.~\eqref{eq:tau_recombination} may be estimated as $\tau_{\mathrm{r}}^{(0)}\sim \tau_{\mathrm{st}}$.

In Eqs.~\eqref{eq:DnuNode}, \eqref{eq:DVpureNode}, and \eqref{eq:sigmaBDsmallT} we neglected impurity scattering,
which broadens the quasiparticle energy levels.  Consequently the result \eqref{eq:sigmaBDsmallT} is valid provided $v_{\mathrm F}\bar{p}_{\mathrm s}, T\gg \Gamma_{\mathrm{el}}$, where $\Gamma_{\mathrm{el}}$ is the characteristic broadening scale of low energy quasiparticle levels.  The value of $\Gamma_{\mathrm{el}}$ is not universal, and depends on the details of the scattering potential. For example,
 for weak impurities
$\Gamma_{\mathrm{el}} \sim \Delta^{2}_{0}\tau_{\mathrm{el}}\exp(-\Delta_{0}\tau_{\mathrm{el}})$ \cite{lee_localized_1993,mineev_introduction_1999}, while in the case of strong impurities  whose scattering cross-section is close to the unitary limit $\Gamma_{\mathrm{el}} \sim \Delta_{0}/\sqrt{\Delta_{0}\tau_{\mathrm{el}}}$, see Refs.~\cite{hirschfeld_resonant_1986,pethick_transport_1986, schmitt-rink_transport_1986}.
In order to  estimate $\sigma_{\mathrm{DB}}$ in the presence of disorder we may evaluate the density of states  using  Eqs.~\eqref{eq:nuDisord} by setting $\tilde{\epsilon} \to  \epsilon + i \Gamma_{\mathrm{el}}$. At relatively large energies,   $\Gamma_{el} < \epsilon < \Delta_0$, the density of states  is practically  unaffected by disorder  and superfluid momentum,
\begin{equation}\label{eq:nu_linear}
  \nu(\epsilon >\Gamma_{\mathrm{el}}, p_\mathrm{s}) \sim \nu_n \frac{\epsilon}{\Delta_{0}}.
\end{equation}
At lower energies, $\epsilon \lesssim \Gamma_{\mathrm{el}}$ it becomes independent of the energy.  In the absence of superfluid current it may be estimated as
\begin{equation}\label{eq:nuBroadening_zero}
  \nu(\epsilon<\Gamma_{\mathrm{el}}, p_{s}=0)\sim \nu_{n}\frac{\Gamma_{\mathrm{el}}}{\Delta_{0}},
\end{equation}
while the correction  to due to the presence of supercurrent, $\delta \nu (\epsilon, p_{\mathrm{s}}) = \nu (\epsilon, p_s) - \nu (\epsilon, 0)$ may be estimated at $v_F p_\mathrm{s} \ll \Gamma_{\mathrm{el}}$ as
\begin{equation}\label{eq:delta_nuBroadening_current}
  \frac{\delta \nu(\epsilon<\Gamma_{\mathrm{el}}, p_{\mathrm s})}{\nu(\epsilon<\Gamma_{\mathrm{el}}, p_{\mathrm s}=0)}\sim
\left(\frac{v_{\mathrm F}p_{\mathrm s}}{\Gamma_{\mathrm{el}}}\right)^{2}.
\end{equation}
Using Eqs.~\eqref{eq:nu_linear}, \eqref{eq:nuBroadening_zero}, and \eqref{eq:delta_nuBroadening_current} we can estimate the level sensitivity $V(\epsilon, p_\mathrm{s}) $ in Eq.~\eqref{eq:level_velocity} as
\begin{equation}\label{eq:level_velocity_estimate}
   V(\epsilon,\bm{p}_{\mathrm s})  \sim
   v_{\mathrm F}\begin{cases}
     \epsilon \frac{v_{\mathrm F} p_{\mathrm s} }{\Gamma^{2}_{\mathrm{el}}} , & \mbox{for } \epsilon < \Gamma_\mathrm{el} \\
     \frac{v_{\mathrm F} p_\mathrm{s}  }{\epsilon}, & \mbox{for} \,\ \epsilon > \Gamma_\mathrm{el} .
   \end{cases}
\end{equation}
Using these estimates, in the temperature interval $v_{\mathrm F}\bar{p}_{\mathrm s}<\Gamma_{\mathrm{el}} <T $   we get
\begin{equation}\label{eq:sigma_intermediate_d}
     \frac{\sigma_{\mathrm{DB}}}{\sigma_{\mathrm D}} \sim \frac{\tau_{\mathrm r}}{\tau_{\mathrm{el}}}\frac{1}{\left[1+(\omega \tau_{\mathrm r})^{2}\right]}\frac{\Delta_0}{T} \left(\frac{v_{\mathrm F} \bar{p}_{\mathrm s}}{\Delta_0} \right)^2  \ln\left(\frac{T}{\Gamma_{\mathrm{el}}} \right).
\end{equation}
We note that at $\Delta_{0}\gg T\gg \Gamma_{\mathrm{el}}$,  the  conductivity  at zero superfluid momentum, $\sigma(\bar{p}_{\mathrm s}=0)\sim \sigma_{\mathrm D}$, is of order the Drude conductivity.  \cite{lee_localized_1993,sun_transport_1995}.

Finally, in  the regime $T, v_{\mathrm F}\bar{p}_{\mathrm s}\ll \Gamma_{\mathrm{el}}$ using Eqs.~\eqref{eq:nuBroadening_zero} and \eqref{eq:delta_nuBroadening_current} we get
\begin{equation}
\frac{\sigma_{\mathrm{DB}}}{\sigma_{\mathrm D}} \sim \frac{\tau_{\mathrm r}}{\tau_{\mathrm{el}}}\frac{1}{\left[1+(\omega \tau_{\mathrm r})^{2}\right]}
\frac{\Gamma_{\mathrm{el}}}{\Delta_{0}}
\left(\frac{T}{\Gamma_{\mathrm{el}}}\right)^{2}\left(\frac{v_{\mathrm F} \bar{p}_{\mathrm s}}{\Gamma_{\mathrm{el}}} \right)^2
\end{equation}

 We note that in this temperature interval $\sigma(\bar{p}_{\mathrm s}=0)\sim \sigma_{\mathrm D} /\Delta_{0}\tau_{\mathrm{el}}\ll \sigma_{\mathrm D}$ \cite{fradkin_critical_1986,lee_localized_1993}.

\section{Discussion \label{sec:conclusion}}

We have shown that supercurrent dependence of the microwave conductivity of superconductors is proportional to the inelastic relaxation time.
Therefore in the presence of supercurrent the absorption coefficient can be larger than the conventional
contribution, which determines the conductivity at $p_{\mathrm s}=0$ and is generally proportional to the elastic mean free time.
We note that such mechanism should exist even in the absence of \emph{dc} supercurrent in  superconductors with broken time-reversal symmetry. For example in topological superconductors with  $p_x+ip_y$ structure of the order parameter where breaking of time reversal symmetry leads to the existence of edge quasiparticle states~\cite{matsumoto_quasiparticle_1999,stone_edge_2004,kallin_chiral_2016}. In time-reversal symmetric superconductors in the absence of  \emph{dc} supercurrent, $\bar{p}_\mathrm{s}=0$, the Debye mechanism of microwave absorption manifests itself
in the anomalously strong non-linear microwave absorption.

The situation with a spatially uniform supercurrent density and electric field, which was considered above, can be realized  in sufficiently  thin superconducting films. In bulk superconductors in the presence of a magnetic field  $ H < H_{c1}$ that is parallel to the surface $\bar{p}_\mathrm{s}$ is nonzero only  within the London  penetration depth  $\lambda_{\mathrm H}$ near the surface.
In this case the situation is different for $s$- and $d$-wave superconductors.

In the $s$-wave case the mechanism of microwave absorption discussed above will still apply to bulk samples and the presented above results still hold up to a numerical factor of order unity.  The reason for this is that the quasiparticles that give the main contribution to microwave absorption have energies that lie in a narrow interval near the gap, $|\epsilon - \Delta| \lesssim \delta \epsilon$, where $\delta \epsilon = v_\mathrm{F} \bar{p}_\mathrm{s}$  in the ballistic regime and $\delta \epsilon = \left( \Delta D^2  \bar{p}_\mathrm{s}^4\right)^{1/3}$ in the diffusive regime.
Roughly half of these quasiparticles have energies below $\Delta$ and therefore they are trapped near the surface within a distance of order $\lambda_{\mathrm H}$.

In bulk samples of gapless $d$-wave superconductors in the presence of a magnetic field parallel to the surface the situation is different. The reason is that the quasiparticles in the relevant energy interval can diffuse into the bulk. Therefore in this case the inelastic relaxation time in corresponding formulas for d-wave superconductors should be substituted by the minimum between the inelastic relaxation time and the time of diffusion from the surface layer of thickness $\lambda_{\mathrm H}$.

Finally we would like to note that the considered above mechanism of the microwave absorption is closely related to the mechanism of \emph{ac} conductivity of SNS junctions discussed in Refs.~\cite{artemenko_theory_1979,zhou_resistance_1997,zhou_density_1998}.

The work of A.A. and M.S. was supported by the U.S.  Department of Energy Office
of Science, Basic Energy Sciences under Award No. DE-FG02-07ER46452 and by the NSF grant  MRSEC  DMR-1719797.
The work of B.S. was funded in part by the Gordon and Betty Moore Foundation's EPiQS Initiative through Grant GBMF4302 and GBMF8686.

\end{document}